\documentclass[journal=jacsat,manuscript=article]{achemso}

\usepackage[version=3]{mhchem} % Formula subscripts using \ce{}
\usepackage{amssymb}
\author{D. Buchta}
\affiliation{Physikalisches Institut, Universit{\"a}t Freiburg, 79104 Freiburg, Germany}
\author{S. R. Krishnan}
\affiliation{Max-Planck-Institut f{\"u}r Kernphysik, 69117 Heidelberg, Germany}
\altaffiliation{Current address: IBM-India Semiconductor R\&D Center, D3-F1, Manyata Park, Bangalore 560045, India}
\author{N. B. Brauer}
\affiliation{Laboratoire de Chimie Physique Mol{\'e}culaire, Swiss Federal Institute of Technology Lausanne (EPFL),1015 Lausanne, Switzerland}
\author{M. Drabbels}
\affiliation{Laboratoire de Chimie Physique Mol{\'e}culaire, Swiss Federal Institute of Technology Lausanne (EPFL),1015 Lausanne, Switzerland}
\author{P. O'Keeffe}
\affiliation{CNR Istituto di Metodologie Inorganiche e dei Plasmi, CP10,00016 Monterotondo Scalo, Italy}
\author{M. Devetta}
\affiliation{CIMAINA, Universit{\`a} degli Studi di Milano, Via Celoria 16, 20133 Milano, Italy}
\author{M. DiFraia}
\affiliation{Elettra-Sincrotrone Trieste, 34149 Basovizza, Trieste, Italy}
\author{C. Callegari}
\affiliation{Elettra-Sincrotrone Trieste, 34149 Basovizza, Trieste, Italy}
\author{R. Richter}
\affiliation{Elettra-Sincrotrone Trieste, 34149 Basovizza, Trieste, Italy}
\author{M. Coreno}
\affiliation{Elettra-Sincrotrone Trieste, 34149 Basovizza, Trieste, Italy}
\author{K. Prince}
\affiliation{Elettra-Sincrotrone Trieste, 34149 Basovizza, Trieste, Italy}
\author{F. Stienkemeier}
\affiliation{Physikalisches Institut, Universit{\"a}t Freiburg, 79104 Freiburg, Germany}
\author{R. Moshammer}
\affiliation{Max-Planck-Institut f{\"u}r Kernphysik, 69117 Heidelberg, Germany}
\author{M. Mudrich}
\email{mudrich@physik.uni-freiburg.de}
\affiliation{Physikalisches Institut, Universit{\"a}t Freiburg, 79104 Freiburg, Germany}

\title[\texttt{achemso} demonstration]
{Penning ionization of doped helium nanodroplets following EUV excitation}

\begin{document}
%%%%%%%%%%%%%%%%%%%%%%%%%%%%%%%%%%%%%%%%%%%%%%%%%%%%%%%%%%%%%%%%%%%%%
%% The manuscript does not need to include \maketitle, which is
%% executed automatically.  The document should begin with an
%% abstract, if appropriate.  If one is given and should not be, the
%% contents will be gobbled.
%%%%%%%%%%%%%%%%%%%%%%%%%%%%%%%%%%%%%%%%%%%%%%%%%%%%%%%%%%%%%%%%%%%%%
\begin{abstract}
Helium nanodroplets are widely used as a cold, weakly interacting matrix for spectroscopy of embedded species. In this work we excite or ionize doped He droplets using synchrotron radiation and study the effect onto the dopant atoms depending on their location inside the droplets (rare gases) or outside at the droplet surface (alkali metals). Using photoelectron-photoion coincidence imaging spectroscopy at variable photon energies (20-25\,eV), we compare the rates of charge-transfer to Penning ionization of the dopants in the two cases. The surprising finding is that alkali metals, in contrast to the rare gases, are efficiently Penning ionized upon excitation of the (n=2)-bands of the host droplets. This indicates rapid migration of the excitation to the droplet surface, followed by relaxation, and eventually energy transfer to the alkali dopants.
\end{abstract}

\section{Introduction}
He nanodroplets are widely used as nanometer-sized cryogenic matrices for spectroscopy of embedded atoms, molecules and clusters~\cite{Toennies:2004,Stienkemeier:2006}. Their peculiar properties such as low temperature (0.38\,K), transparency for visible and UV light, their ability to efficiently cool embedded species (`dopants'), their chemical inertness with respect to dopant-He interaction, and the high mobility of dopants inside the droplets, make them a nearly ideal spectroscopic matrix. However, upon irradiation with extreme ultraviolet (EUV) light, where He droplets are strongly absorbing, a complex photo dynamics is initiated by the excitation or ionization of He atoms inside the droplets~\cite{Froechtenicht:1996,Joppien:1993,Haeften:1997,Peterka:2003,Kornilov:2011,BuenermannJCP:2012}.

Three regimes of excitation and ionization can be distinguished:

(i) At photon energies $20.5<h\nu<23\,$eV, He nanodroplets are excited with high cross sections into perturbed excited states (``bands'') correlating to the 1s2s$^1$S and 1s2p$^1$P He atomic levels. Fast droplet-induced intra-band and inter-band relaxation as well as He$_2^*$ excimer formation follows the excitation~\cite{Kornilov:2011,Buenermann:2012}. Due to the repulsive interaction between excited He$^*$ or He$_2^*$ and the He environment the excitation is presumed to migrate to the surface by a resonant hopping process or by fast atomic motion within 10-20\,ps~\cite{Buchenau:1991,Scheidemann:1993,Buenermann:2012}. Depending on the size of the He droplet, the He$^*$(1s2p$^1$P) state is emitted into vacuum or trapped at the surface and eventually relaxes into the long-lived 1s2s$^{1,3}$S states or into vibrationally excited He$_2^*$ molecules. The latter are subject to vibrational relaxation by coupling to the He droplet and eventually evaporate off the droplet surface~\cite{Buchenau:1991}.
%Note that long-lived metastable excitations in bulk liquid helium (He$_2^*$ in the $a^3\Sigma_u^+$-state) are known to be trapped in bubbles which propagate over macroscopic distances to the He surface where they are weakly bound~\cite{Buchenau:1991}.

(ii) At photon energies $23<h\nu<24.6\,$eV, the droplet response is even more complex. In addition to the aforementioned relaxation channels, the emission of He$^*$ and He$_2^*$ in Rydberg states becomes dominant~\cite{Buenermann:2012,BuenermannJCP:2012}, and the fraction of molecules increases with rising excitation energies~\cite{Haeften:1997}. At $h\nu>24\,$eV the population of triplet states of He was also observed~\cite{Haeften:1997,Kornilov:2010}. As a further relaxation channel, autoionization of He droplets sets in at $h\nu>23\,$eV leading to the formation of small ionic fragments (He$_n^+$, $n\leq 17$) as well as large cluster ions ($N\gtrsim 10^3$)~\cite{Froechtenicht:1996}. A peculiarity of the ionization of He droplets below the ionization energy $E_{i,\mathrm{He}}=24.6\,$eV of atomic He is the emission of electrons with very low kinetic energy $<1\,$meV as seen in photoelectron imaging experiments~\cite{Peterka:2003,Peterka:2007}. Recent time-resolved photoelectron and photoion imaging experiments using femtosecond light pulses in the EUV from high-order harmonic generation have revealed the dynamics of various relaxation processes~\cite{Kornilov:2010,Kornilov:2011,Buenermann:2012,BuenermannJCP:2012}. In particular the impact on the relaxation dynamics of the location of the excitation was shown, in line with recent time-resolved fluorescence measurements~\cite{Haeften:2005,Haeften:2011}. While the broad and blue shifted 1s2s$^1$S and 1s2p$^1$P-bands around $h\nu=20.8$ and $21.6\,$eV were attributed to excitations of the bulk region of the droplets, the higher excitations at $h\nu>23\,$eV are assumed to be predominantly located at the droplet surface.

(iii) At photon energies $h\nu>E_{i,\mathrm{He}}$, He$^+$ ions (positive holes) are created inside or at the surface of the droplets. Due to fast resonant hopping the positive charge migrates towards the droplet center on a time scale of 60-80\,fs before being trapped by forming a He$_2^+$ molecular ion~\cite{Stace:1988,Scheidemann:1993,Halberstadt:1998,Seong:1998}. The internal energy of the molecule as well as the binding energy liberated upon formation of `snowball' structures (He atoms tightly bound around the ion core) is believed to stop the charge-hopping process by causing massive fragmentation of the droplet. Therefore, He$^+$ largely from background He atoms and He$_2^+$ from droplets are the dominant species appearing in the mass spectra~\cite{Froechtenicht:1996,Peterka:2007,Kim:2006}.

Only a few synchrotron studies of doped He nanodroplets have been reported to date~\cite{Froechtenicht:1996,Peterka:2006,Kim:2006,Wang:2008}. These studies are restricted to dopants that are immersed in the He droplet interior. However, alkali metal atoms and molecules which are located in dimple-like states at the droplet surface, have been studied using electron impact ionization~\cite{Scheidemann:1997,Lan:2011}. In the latter experiments a clear onset of alkali ion formation was observed at electron energies below 23\,eV, that is, in regime (i) of droplet excitation. This was rationalized by excitation transfer (akin to Penning ionization in binary collisions) being the primary ionization mechanism in the case of alkali adducts. In contrast to the positive He$^+$ hole which migrates into the droplet interior, He$^*$ excitations tend to be repelled outwards to the droplet surface where the alkali dopants are located. Note, however, that electron bombardment is likely to produce metastable He 1s2p$^3$P and 1s2s$^3$S states which are not directly accessible by optical excitation.

In the previous synchrotron experiments, fragment ion mass spectra as well as photoelectron spectra were recorded. In all studies, the dopant ion signal or the correlated electron signal were entirely generated by indirect ionization due to charge or excitation transfer from the He surrounding the dopant~\cite{Peterka:2006,Kim:2006,Wang:2008}. The highest dopant ion yield was found in regimes (iii) and (ii) where the He droplets are directly ionized or autoionize, respectively. As for electron impact, the initially formed He$^+$ charges migrate through the droplets towards the dopant, steered by electrostatic ion-dopant polarization forces. For droplets of sizes $N=200$-$15000$ He atoms, the charge-transfer probabilities vary from 80 to 5$\,\%$, respectively~\cite{Callicoatt:1996,Callicoatt:1998,Ruchti:2000,Lewis:2005,Peterka:2006}.

When exciting the droplets into the 1s2s$^1$S and 1s2p$^1$P bands (regime (i)), weak dopant ionization rates were observed due to excitation transfer (Penning) ionization~\cite{Froechtenicht:1996,Peterka:2006,Kim:2006,Wang:2008}. Photoelectron spectra of He nanodroplets doped with rare-gas atoms have revealed that Penning ionization is likely to proceed in a two-step process where electronic relaxation from He(1s2p$^1$P) into He(1s2s$^1$S) precedes the excitation transfer step to the dopant~\cite{Wang:2008}. The photoelectron angular distributions produced by dopant Penning ionization were found to be markedly more isotropic than for gas-phase atoms. The photoelectron spectra correlated to Penning ionization of SF$_6$, however, showed only minor differences when compared to those of free SF$_6$, pointing at optical-like  electronic dipole interaction to be active rather than ``traditional'' Penning ionization in collisions involving metastable atoms~\cite{Peterka:2006}.

In the present work we study EUV ionization of He nanodroplets doped with alkali atoms (Li, Na, K) in comparison with rare gas (Ar) atoms by means of photoelectron-photoion coincidence (PEPICO) imaging spectroscopy. The experiments were carried out at the GasPhase beamline at the synchrotron Elettra, Trieste, Italy. They complement earlier work of the group of D. Neumark who studied pure and doped He droplets using synchrotron radiation~\cite{Peterka:2003,Peterka:2006,Peterka:2007,Wang:2008,Kim:2006} and more recently using EUV light generated by high-harmonics of intense ultrashort laser pulses~\cite{Kornilov:2011,Buenermann:2012}. While our results for rare-gas doped He droplets corroborate the previous studies~\cite{Wang:2008,Kim:2006}, we find significantly different ionization dynamics for alkali-doped He droplets. While charge transfer ionization is the dominant channel of indirect ionization of embedded atoms and molecules, we find a strongly enhanced probability of ionizing alkali metals by a Penning process involving He and He$_2$ in various excited states. We attribute this observation to the migration of He excitations to the He droplet surface where the alkali dopants are located. Thus, our results add complementary information to the current discussion of the location and dynamics of excited states within the bulk and surface regions of He nanodroplets~\cite{Closser:2010,Haeften:2011,Kornilov:2011,Buenermann:2012}. Our findings are discussed in the context of various relaxation and ionization mechanisms inferred from previous fluorescence, ion mass spectrometric and photoelectron studies.

\begin{figure}[H]
\centering
\includegraphics[width=0.9\textwidth]{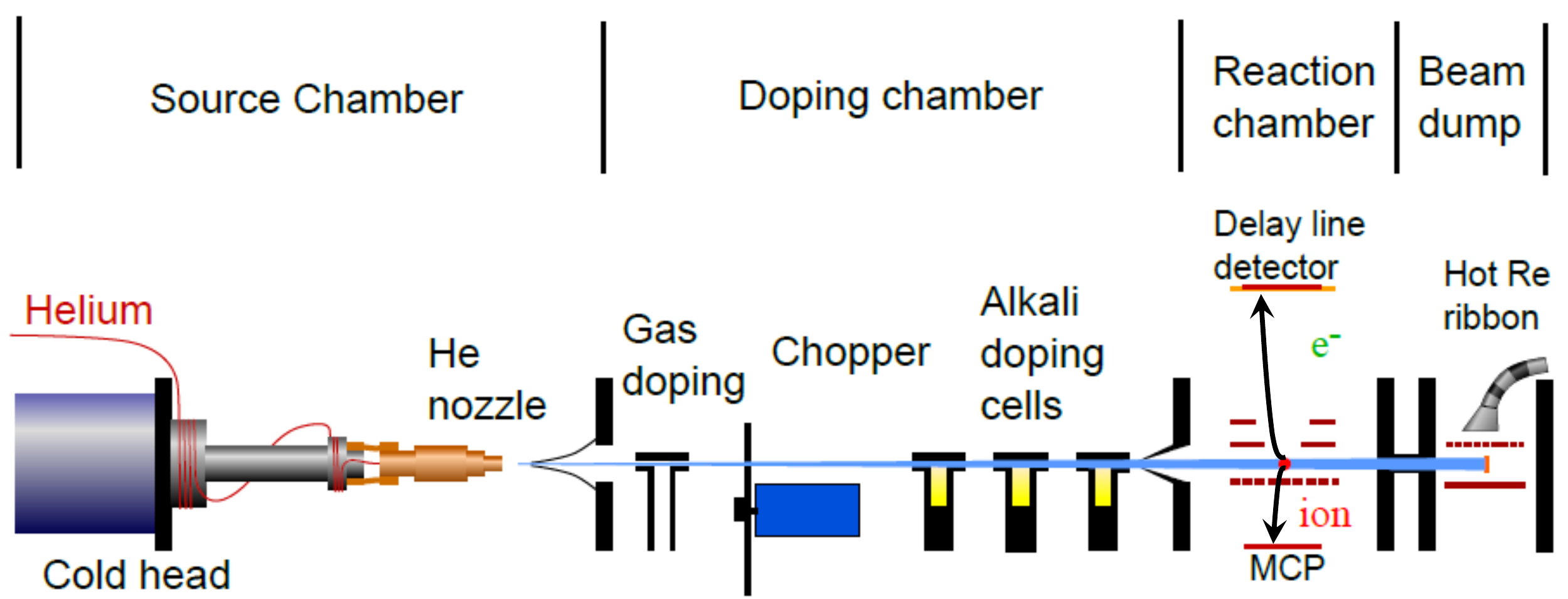}
\caption{Sketch of the experimental setup. The left part depicts the cryogenic source of He nanodroplets, the center part represents gas and vapor cells for doping the droplets with rare gases and with alkali metals, respectively, and the right part sketches the electron and ion detector as well as a surface ionization detector used for monitoring the alkali doping level.}
\label{fig:setup}
\end{figure}
\section{Experimental}
The experiments presented here are performed using a mobile He droplet machine attached to an imaging photoelectron-photoion coincidence (PEPICO) detector at the GasPhase beamline of the synchrotron Elettra, Trieste, Italy~\cite{OKeeffe:2011}. The He droplet beam apparatus is sketched in Fig.~\ref{fig:setup} and resembles previously used setups~\cite{Krishnan:2012,Fechner:2012}. In short, a beam of He nanodroplets is produced by continuously expanding pressurized He (50\,bar) of high purity (He 6.0) out of a cold nozzle ($T_0=13$-$33\,$K) with a diameter of $5\,\mu$m into vacuum. At these expansion conditions, the mean droplet sizes can be inferred from previous experiments to range between 200 and 17000 He atoms per droplet~\cite{Toennies:2004,Stienkemeier:2006}.

After passing a skimmer ($0.4\,$mm) the He droplets are doped with either rare gas atoms in a scattering cell with a length of 30\,mm or with alkali atoms (Li, Na, K) in one of three heated cells with a length of 10\,mm each. The doping level of rare gases is adjusted using a dosing valve and monitored by measuring the pressure increase in the doping chamber. The alkali cells are heated to $400\,^\circ$C for Li, $200\,^\circ$C for Na and $125\,^\circ$C for K which corresponds to vapor pressure values around the level required for maximum likelihood of single atom doping. A mechanical beam chopper is used in all measurements to discriminate any signals correlating with the droplet beam from background signals related to He, rare gases or alkali atoms effusing into the detection region. The He droplet beam intensity as well as the alkali doping level is monitored using a beam dump chamber attached to the end of the apparatus which contains a simple surface ionization detector~\cite{Stienkemeier:2000}.

In the detector chamber, the doped He droplet beam crosses the synchrotron beam at right angle in the center of an electrode arrangement that accelerates photoelectrons onto a position and time resolving delay-line detector, and photoions onto a microchannel plate detector that records flight times. The measurement of electrons and ions in coincidence allows us to extract from the data both ion mass spectra and mass correlated velocity-map photoelectron images. The latter are transformed into photoelectron spectra as well as angular distributions using standard Abel inversion programs~\cite{Vrakking:2001,Garcia:2004}. Depending on the voltage applied to the electrodes, electrons with kinetic energies up to 30\,eV can be detected.

The synchrotron radiation at the GasPhase beam line exits a U12.5 undulator and passes a variable angle spherical gratings monochromator. The photon energy $h\nu$ can be varied on demand between 13 and 900\,eV. In the measurements presented here $h\nu$ is restricted to the range of electronic excitations of He atoms and droplets, up to a few eV above the first ionization threshold (20-26\,eV). The intensity of the radiation is monitored by a photodiode and all photon energy dependent ion and electron spectra shown in this work are normalized to this intensity signal. Note that a non-negligible amount of second order radiation is present at the lower end of the tuning range $h\nu\lesssim 20\,$eV. The energy resolution $E/\Delta E$ is $\gtrsim 10^4$. The pulse repetition rate is 500\,MHz and the peak intensity in the interaction region is estimated to range around $\sim 15\,$W\,m$^{-2}$.

\section{Results and discussion}
\subsection{Mass spectra}
In this work we focus on the comparison of photoion and photoelectron signals measured for He nanodroplets doped with Ar as a typical representative of rare gas dopants with ionization signals obtained from alkali-doped (Li, Na, K) droplets at various photon energies in the range $h\nu=20$-$26$\,eV.

\begin{figure}[H]
\centering
\includegraphics[width=1.0\textwidth]{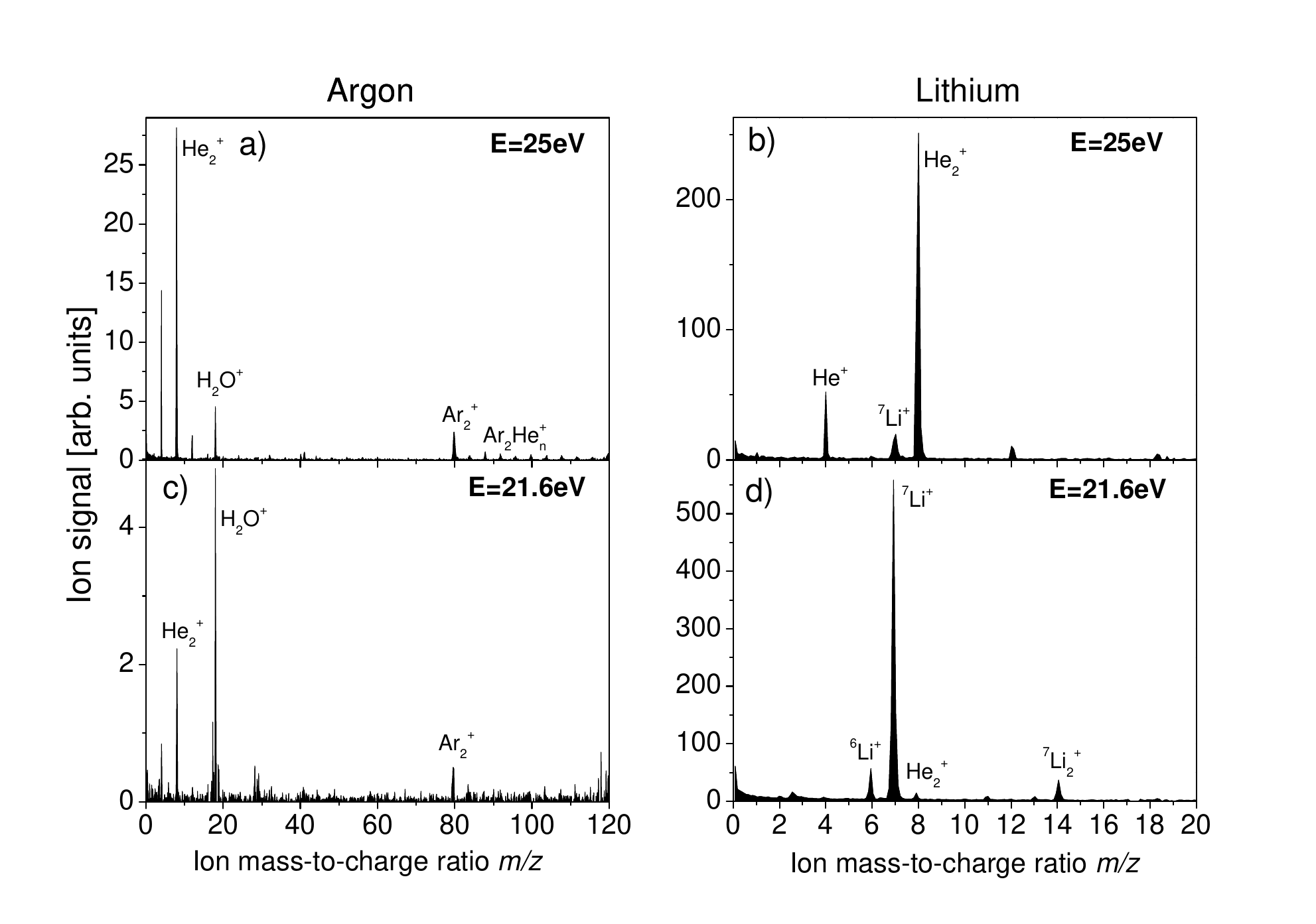}
\caption{Difference mass spectra of Ar and Li doped helium nanodroplets at photon energies $h\nu=21.6\,$eV and $h\nu=25\,$eV. The He expansion conditions are $p_0=50\,$bar and $T_{0}=23\,$K ($N=1900$) for the spectra in a) and c) and $T_0=19\,$K ($N=4500$) for the spectra in b) and d).}
\label{mass_spectra}
\end{figure}

Let us start the discussion of experimental results with the presentation of typical ion mass spectra. The dependence of the dopant mass signals on the experimental parameters (photon energy, He droplet size) will be discussed subsequently. Fig.~\ref{mass_spectra} compares the mass spectra recorded at $h\nu=21.6\,$eV, which corresponds to the maximum of the 1s2s$^1$S$\rightarrow$1s2p$^1$P absorption band of He nanodroplets~\cite{Froechtenicht:1996,Joppien:1993,Haeften:2001} to those recorded at $h\nu=25\,$eV, where the He atoms in the droplets are directly ionized. The He droplet beam source is operated at a He pressure $p_0=50\,$bar and at nozzle temperatures $T_0=23\,$K when doping with argon and $T_0=19\,$K when doping with lithium. The corresponding  mean He droplet sizes amount to $N=1900$ and $4500$, respectively~\cite{Toennies:2004}. The mass spectra in Fig.~\ref{mass_spectra} a) and c) are recorded when doping the He droplets with on average $n_{Ar}\approx 0.8$ Ar atoms, those in b) and d) are measured when doping with $n_{Li}\approx 0.4$ Li atoms. The use of a mechanical He droplet beam chopper allows to discriminate the ion signals originating from the doped He droplet beam from background gas ions. When the beam chopper is in the `open' position, both contributions are measured whereas in the `closed' position, only background ions contribute. Thus, the shown difference signal gives the He droplet-correlated contribution.
%Unfortunately, this discrimination is not perfect so that a cross talk between the channels remained, as can be seen when comparing chopper on/off signals at the masses of He$^+_n$ or Li$_2^+$.

At photon energies exceeding $E_{i,\mathrm{He}}$, the dominant mass peaks in the spectra besides residual gas components (H$_2$O) are those of He$^+$ and He$_2^+$ (Fig.~\ref{mass_spectra} a) and b)). Note that He$_2^+$ is even more abundant than He$^+$, in contrast to earlier electron impact and synchrotron experiments~\cite{Froechtenicht:1996,Buchenau:1991,Peterka:2006,Kim:2006}. This may be due to the long flight distance from the nozzle up to the ionization region of $71\,$cm in our experiment, which results in a highly collimated droplet beam where the content of free He atoms is suppressed. The efficient formation of He$_2^+$ ions agrees with the established notion that the initially created He$^+$ positive hole migrates within the He droplets before localizing by forming a He$_2^+$ ion. The binding energy liberated by forming the He$_2^+$ molecule as well as by forming a tightly bound shell of He atoms around the ion (`snowball') subsequently induces droplet fragmentation and the ejection of bare He$_2^+$. Higher He$_n^+$ cluster ion masses are also present with lower intensities in the entire shown mass range when the He droplet beam is on. In the case of doping with Ar (Fig.~\ref{mass_spectra} a)), a prominent Ar$_2^+$ mass peak is visible at $m/z=80\,$u. This signal reflects charge transfer ionization of Ar dopants that have aggregated to form Ar$_2$ dimers inside the He droplets~\cite{Scheidemann:1990,Scheidemann:1993,Callicoatt:1998,Kim:2006}. This indirect ionization process was found to be efficient with a probability of charge exchange from a He$^+$ ion within the droplet ranging between 5 and 40\% depending on the droplet size. The reduced detection efficiency of the atomic Ar$^+$ ion was attributed to the high probability of Ar$^+$ to remain bound to the He droplet instead of being ejected as free ions~\cite{Callicoatt:1998}. In addition, we find small He$_{n}$Ar$_{2}^{+}$ complexes as in electron-impact experiments~\cite{Callicoatt:1998}. When doping with Li (Fig.~\ref{mass_spectra} b)), a small mass peak corresponding to $^7$Li$^+$ is visible at $h\nu=25\,$eV.
%The nearly constant signal background in the mass spectra with chopper-open versus chopper closed is presumably due to photons scattered from the He droplets, either by resonant excitation followed by spontaneous emission or by Rayleigh-scattering~\cite{Peterka:2006}.

The mass spectra significantly change when the photon energy is tuned to $h\nu=21.6\,$eV (Fig.~\ref{mass_spectra} c) and d)). Since this energy falls below the autoionization threshold of He droplets we do not expect to detect He$^+$ ion signals. The presence of He$^+$ mainly from background He gas as well as of He$_2^+$ from the droplet beam is due to direct ionization by the higher order content of the synchrotron radiation mostly at $2\times h\nu=43.2\,$eV. This interpretation is confirmed by analyzing the photoelectrons correlating to the He$^+$ and He$_2^+$ ions. The photoelectron spectra are peaked at $18.8\,$ eV as expected from the energy balance $2\times h\nu-E_{i,\mathrm{He}}=18.6\,$eV. Accordingly, the He droplet-correlated Ar$_2^+$ signal results mainly from charge transfer ionization by the directly ionized He. Ionization of Ar by the transfer of excitation in a Penning-type process as studied in Ref.~\cite{Wang:2008} is likely to contribute to the signal as discussed below. The differences in the ionization spectra are even more visible in the case of Li doping (Fig.~\ref{mass_spectra} d)). The Li$^+$ ion peak is enhanced by about a factor 20 compared to the spectrum at $25\,$eV yielding the highest ion count rate overall. In addition, the less abundant isotope $^6$Li$^+$ as well as Li$_2^+$ become apparent as expected from the Poissonian pick-up statistics. This indicates a dramatically enhanced Penning ionization probability for Li as compared to Ar.

\begin{figure}[H]
\centering
\includegraphics[width=0.6\textwidth]{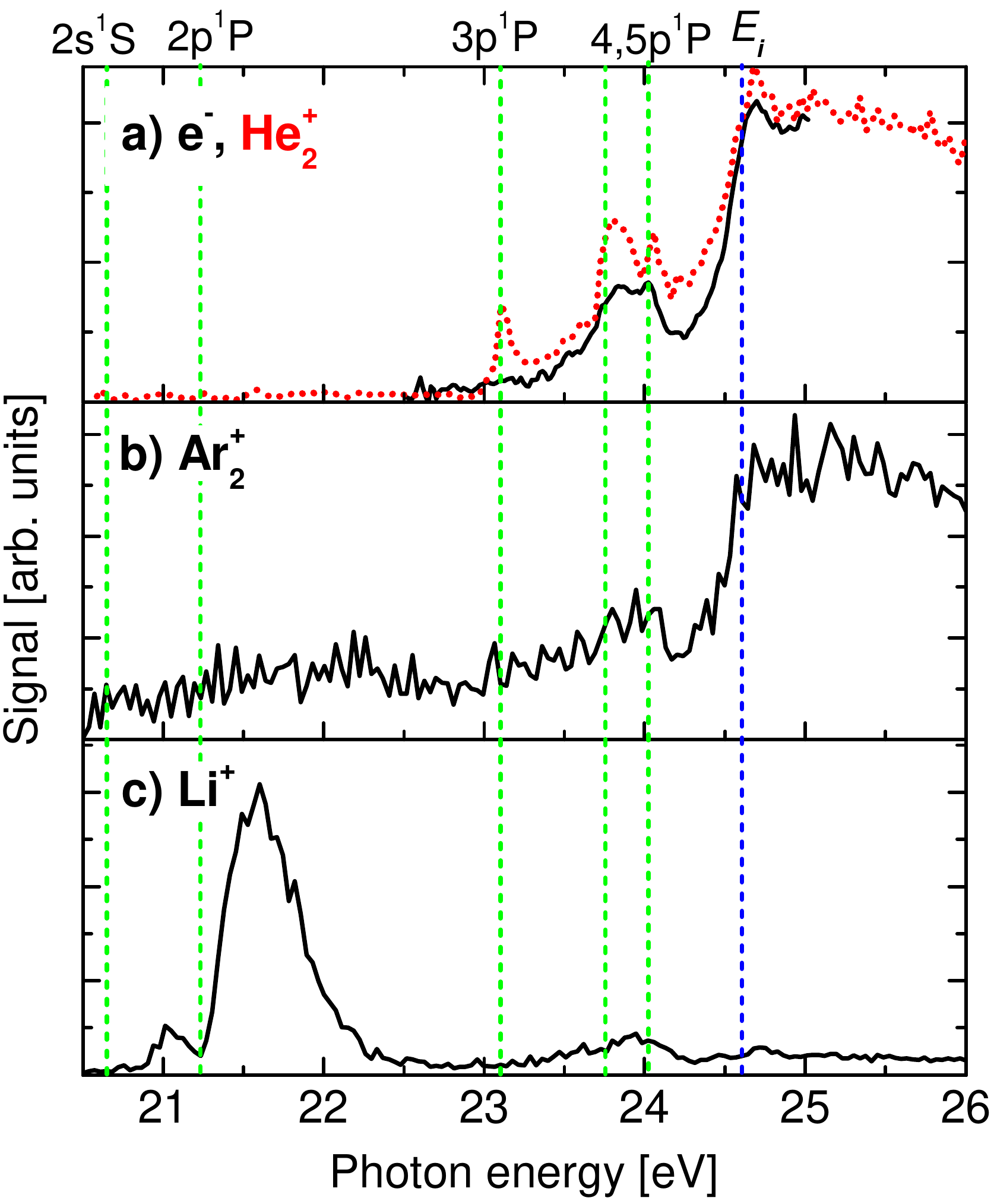}
\caption{Photon energy dependence of the yield of photoelectrons and He$_2^+$ ions (a), Ar$_2^+$ (b) and Li$^+$ ions (c) measured by illuminating pure (a) and doped He nanodroplets (b), (c). The vertical dashed lines indicate He atomic level energies.
%The He expansion conditions are $p_0=50\,$bar and $T_0=21\,$K for the spectrum a), 23\,K for b) and 19\,K for c).
}
\label{fig:Scans_1}
\end{figure}
\subsection{Synchrotron spectra}
The dependence of the He droplet mediated ionization efficiencies of dopants on the photon energy $h\nu$ is studied by recording the electron and dopant ion signals while varying $h\nu$. The resulting spectra are depicted in Fig.~\ref{fig:Scans_1}. While the He expansion pressure is held constant at $50$\,bar, the He nozzle temperature is set to $T_0=21\,$K corresponding to a mean droplet size $N=2900$ for the measurement shown in a) and to $T_0=23\,$K ($N=1900$) and $T_0=19\,$K ($N=4500$) for those in b) and c), respectively. At these temperature values we measure the highest absolute dopant ion signal rates. The total electron (solid line) and He$_2^+$ ions signals (dotted line) shown in Fig.~\ref{fig:Scans_1} a) using different vertical scales are measured with undoped neat He droplets.
% in the range $22.5\leq h\nu\leq 25\,$eV. Below this range no He droplet-correlated electrons are observed.
The broad band structure at photon energies $23\leq h\nu\leq 24.6\,$eV is in good agreement with previous ionization spectra recorded with neat He nanodroplets~\cite{Froechtenicht:1996}. Note that the corresponding spectra by recording the total ion yield or the He$_2^+$ yield closely resemble the electron spectrum. However, we systematically measure higher electron count rates than total ion yield by a factor of 5-15 depending on the doping conditions and on $h\nu$. This indicates the formation of large He$_n^+$ cluster ions with $n>100$ that fall beyond the detection range of our setup~\cite{Froechtenicht:1996}. The peaked structures around $21.8$, $23.1$, $23.8$ and $24.7\,$eV can be assigned to excited He droplet states that mostly correlate to the 1s2p$^1$P, 1s3p$^1$P, 1s4p$^1$P, and highly excited Rydberg levels of atomic He. All excited droplet states except those belonging to the 1s2s, 1s2p configurations are subject to vibrational autoionization yielding mostly He$_2^+$ ions and electrons with extremely low kinetic energies~\cite{Peterka:2003,Peterka:2007,Kornilov:2010,Kornilov:2011}.

\begin{figure}[H]
\centering
\includegraphics[width=0.55\textwidth]{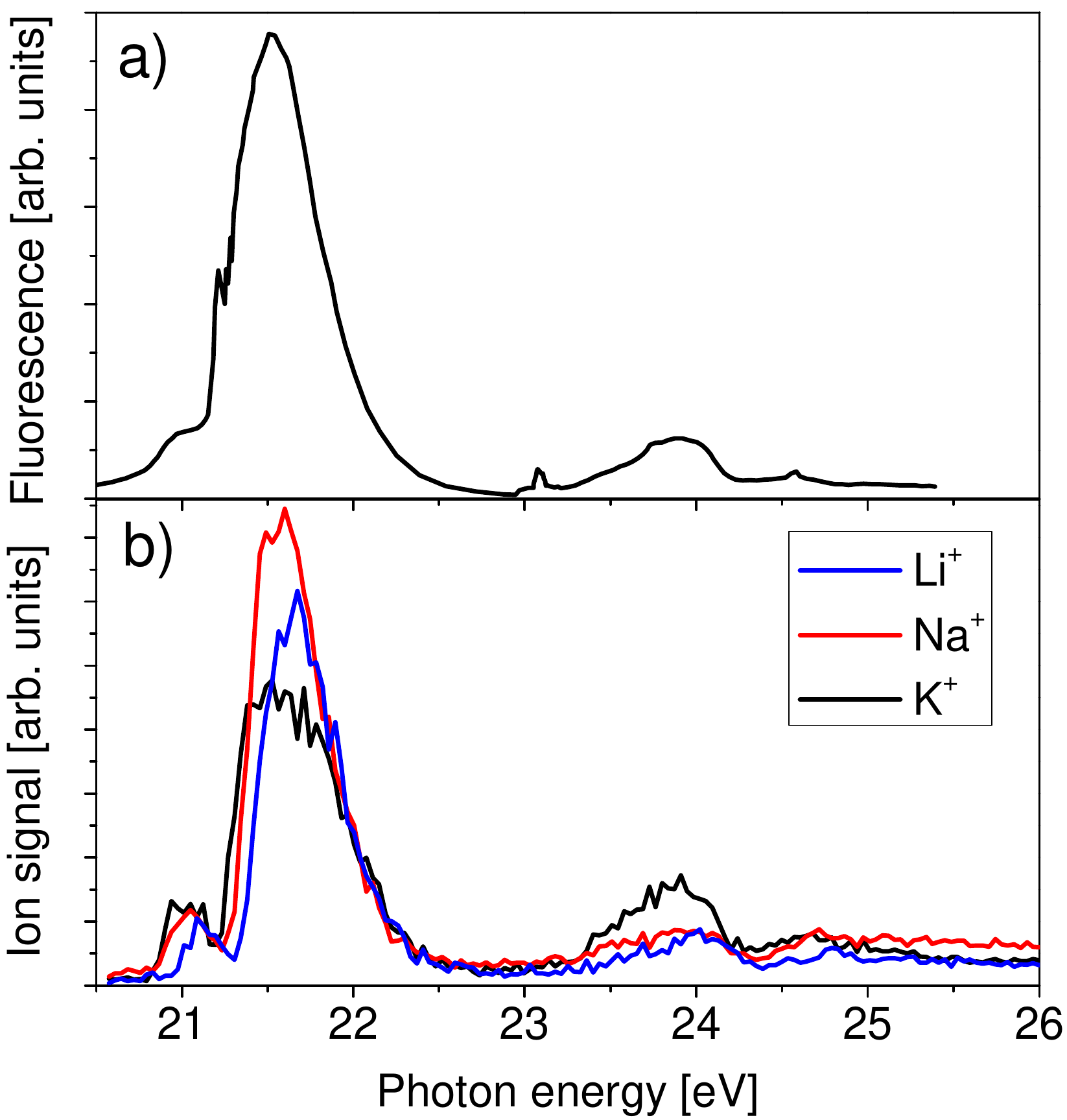}
\caption{Comparison of photoionization spectra recorded at the alkali dopant masses b) with the fluorescence measurements of Joppien \textit{et al.}~\cite{Joppien:1993} a).
The He expansion conditions are $p_0=50\,$bar and $T_{0}=19\,$K ($N=4500$).}
\label{fig:Scans_alkali}
\end{figure}
The ionization spectra of Ar and Li dopants are shown in Fig.~\ref{fig:Scans_1} b) and c). The conspicuous difference between the two traces is the main result of this paper. The Ar$_2^+$ spectrum closely follows the one of He$_2^+$ yielding highest count rates at $h\nu>E_{i,\mathrm{He}}$. While the signal-to-noise ratio is inferior to the electron and He$_2^+$ data, the band structure measured in the electron spectrum with a maximum around 24\,eV and a sharp increase for $h\nu>24.3\,$eV is clearly recovered in the Ar$_2^+$ spectrum. In addition, a weakly resolved broad maximum in the range of the 1s2p$^1$P band around 21.6\,eV indicates some contribution to the Ar$_2^+$ signal by Penning ionization. This peak acquires slightly higher contrast when the mean droplet size is reduced to $N=200$ but never nearly surpasses the height of the signal at $h\nu\geq 25\,$eV as instead observed for Li.

The Li$^+$ spectrum is strongly peaked at the 1s2p$^1$P band around 21.6\,eV. The side band at 21.2\,eV coincides with the absorption band of He droplets associated with the atomic 1s2s$^1$S level~\cite{Joppien:1993,Haeften:2001}. Surprisingly, the Li$^+$ signal produced by charge transfer ionization at $h\nu>25\,$eV stays behind the one due to Penning ionization from any excited He droplet state except the 1s3p$^1$P level. An earlier study of electron impact excitation and ionization of He droplets doped with Li showed an appearance threshold of Li$^+$ ions at about 19\,eV, significantly lower than the one for He$^+$ and He$_2^+$~\cite{Scheidemann:1997}. While the low energy resolution did not allow the authors to resolve the individual droplet bands associates with the Li$^+$ signal, they inferred that the metastable 1s2s$^3$S and 1s2s$^1$S states caused the Li$^+$ signal onset at low energies. In contrast, our measurements unravel the contribution of the He droplets bands to the Penning ionization signal with high spectral resolution. In particular, the high Li$^+$ signal in the 1s2p$^1$P band shows that Penning ionization of surface-bound dopants is as efficient when exciting short-lived ($\lesssim 550\,$ps) states~\cite{Drake:2006} as for metastable states.

The interpretation of the ionization of Li in terms of a Penning process via excited states of He droplets is nicely confirmed by the comparison with the fluorescence excitation spectrum recorded by Joppien et al.~\cite{Joppien:1993} shown in Fig.~\ref{fig:Scans_alkali}. The ionization traces of all studied alkali species Li, Na, K closely follow the fluorescence spectrum. Only in the case of K the band around 24\,eV is slightly more pronounced. Thus, the ionization of alkali adducts sensitively probes the cross section for excitation of He droplets, nearly irrespective of the final state of excitation. One may argue that the alkali dopants are preferentially ionized by transfer from He$^*$ created in the outer surface region close to the dopants in the first place. However, the absorption spectrum of He atoms in that region of low density resembles more the one of free He atoms with narrow peaks close to 21.2\,eV and 23.1\,eV which are visible in the fluorescence spectrum~\cite{Haeften:2005}. Since our alkali ionization spectra only reproduce the broadband structure associated with He$^*$ excitations in the bulk of the droplets~\cite{Haeften:2005} we argue that while He$^*$ are mostly created in the droplet interior, they migrate towards the alkali dopant at the surface driven by repulsive short-range He-He$^*$ forces and possibly by alkali-He$^*$ polarization forces. Note that also the peak shapes of Penning and fluorescence signals are very similar. Thus, Penning ionization is equally efficient for the excitation of the lower and the higher energetic edges of each band, which are commonly attributed to localized excitations at the surface or in the bulk, respectively~\cite{Closser:2010,Kornilov:2011}. This may indicate that homogeneous peak broadening actually dominates over the assumed inhomogeneous broadening effect obtained by averaging over shifted atomic levels due to varying He density in different regions of the droplet.

%(*Mention the $K_n$ measurements*)

In order to assess the efficiency of the He mediated ionization of alkalis via the Penning process we tried to measure the yield of alkali ions from doped He droplets obtained by direct photoionization at photon energies $h\nu<20\,$eV, that is, below the lowest excitations of He droplets. The intensity of higher order radiation was reduced using a neon (Ne) gas filter installed in the beamline connecting the detector chamber to the synchrotron. However, even at the highest possible Ne gas pressure, direct ionization of He could not be sufficiently reduced to suppress charge transfer ionization of the alkali dopants. Thus, for a typical He droplet size $N=2000$ we can only specify a rough upper bound for the ratio of He droplet induced Penning ionization rate with respect to direct ionization of $>10^{3}$. From an estimated absorption cross section of He atoms in He droplets of $\sigma_{a,He}=25\,$Mb and a photoionization cross section of Li of $\sigma_{i,Li}=0.77\,$Mb~\cite{Hollauer:1990} we conclude that the Penning process is efficient to at least 1\%.

\begin{figure}[H]
\centering
\includegraphics[width=0.6\textwidth]{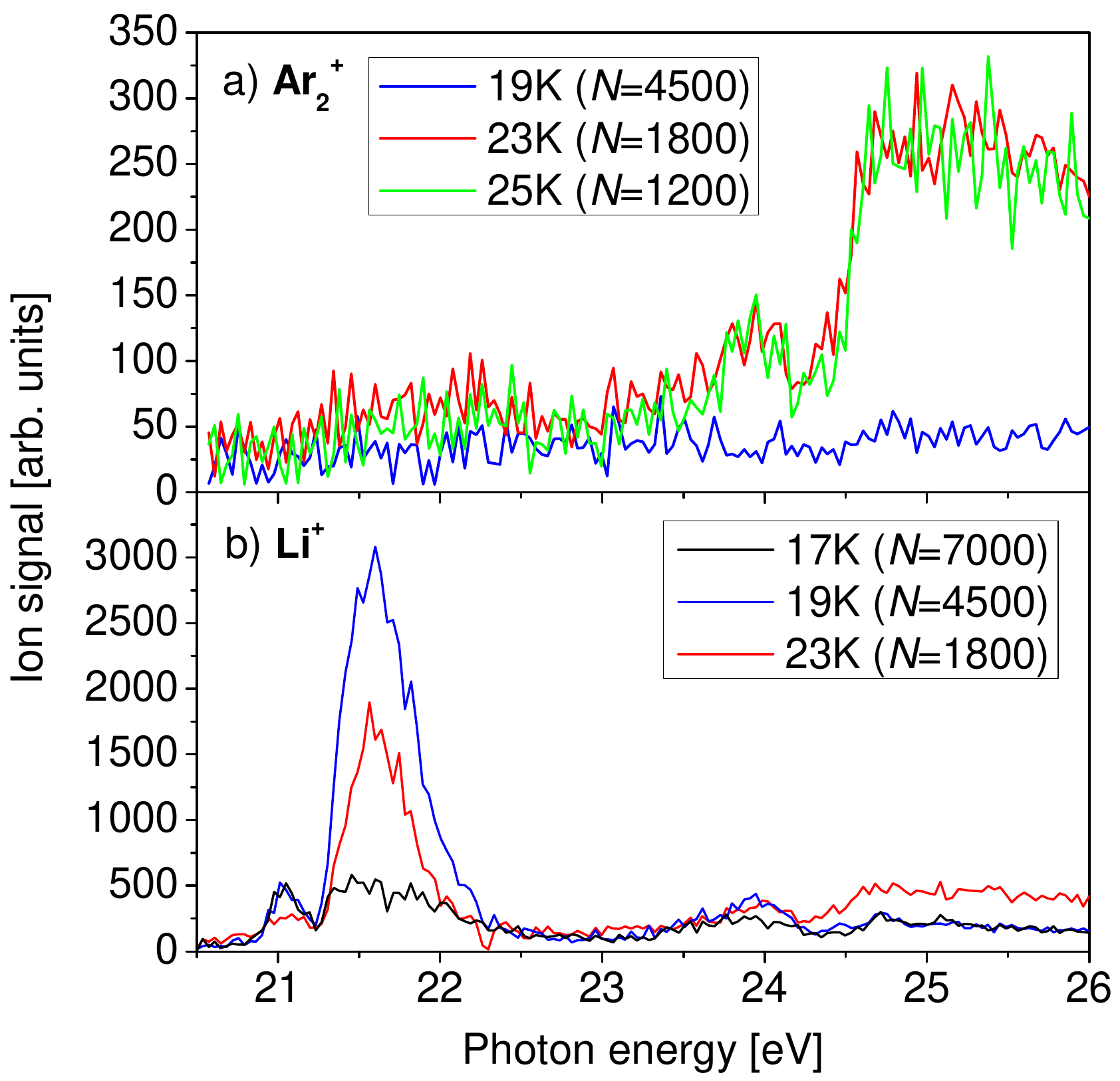}
\caption{Li$^{+}$ (a) and Ar$_{2}^{+}$ (b) dopant ion abundance spectra as a function of the photon energy for various nozzle temperatures $T_0$ corresponding to He droplet sizes $N$.}
\label{fig:nozzlescans}
\end{figure}
The process of charge transfer ionization of dopants following resonant hopping of an initially created He$^+$ positive hole has been previously studied experimentally for varying mean droplet sizes $N$~\cite{Callicoatt:1996,Callicoatt:1998,Ruchti:2000,Lewis:2004}. Since charge transfer ionization of the dopant is assumed to compete with the formation of He$_2^+$, the charge transfer probability increases when reducing $N$ and thereby reducing the distance until He$^+$ reaches the dopant, in agreement with simulations~\cite{Stace:1988,Halberstadt:1998,Seong:1998,Lewis:2004}.

\begin{figure}[H]
\centering
\includegraphics[width=0.6\textwidth]{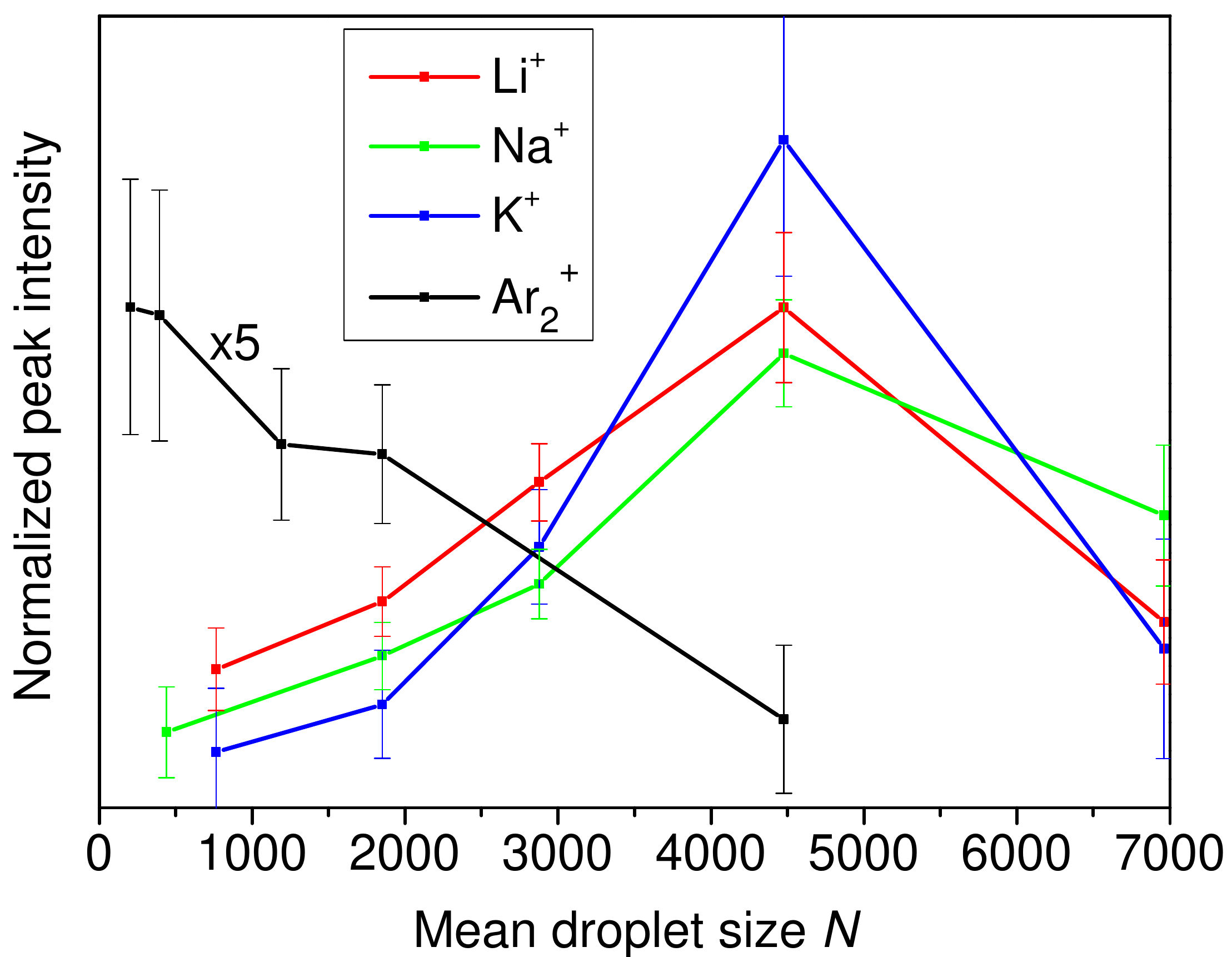}
\caption{Integrated Penning (Li$^+$, Na$^+$, K$^+$, 21.2-22.2\,eV) and charge transfer ionization (Ar$_2^+$, 25-26\,eV) signals as a function of mean droplet size. All values are normalized to the corresponding He$_2^+$ signals.}
\label{fig:peakintensity_nozzle}
\end{figure}
The droplet size dependencies of Ar$_{2}^{+}$ and Li$^{+}$ ionization spectra are shown in Fig.~\ref{fig:nozzlescans} a) and b).
%Significant Ar$_{2}^{+}$ count rates only appear at nozzle temperatures $T_0\gtrsim 20\,$K.
At nozzle temperatures in the range $T_0=23$ ($N=1900$) up to $33\,$K ($N=200$) the Ar$_{2}^{+}$ spectrum is approximately unchanged and then drops in overall amplitude without changing its structure when $T_0$ is further reduced to $T_0=19\,$K ($N=4500$). In contrast, the prominent Li$^+$ peak around 21.6\,eV strongly varies in amplitude (Fig.~\ref{fig:nozzlescans} b)). It is highest at $T_0=19\,$K, both in absolute terms as well as in proportion to the charge transfer ionization signal at $h\nu>25\,$eV. A more quantitative comparison of the characteristic Ar$_{2}^{+}$ and Li$^{+}$ rates is provided by Fig.~\ref{fig:peakintensity_nozzle} which shows the integrated signals normalized to the He$_2^+$ count rate. Since the latter makes up more than $65\%$ of the total ion yield coming from the He droplets, the shown ratio of Ar$_{2}^{+}$ and Li$^{+}$ to He$_2^+$ count rates can be associated with the probability for indirect ionization via ionized or excited He droplets, respectively. Note that the ratio of Li$_2^{+}$ to Li$^{+}$ ion counts nearly remains constant in this range of droplet size variation so that a significant influence of changing doping statistics due to varying pick-up cross sections of the droplets can be excluded. The range of integration (25-26\,eV) for Ar$_2^+$ corresponds to direct He ionization whereas the integration interval 21.2-22.2\,eV used for Li$^+$ covers the 1s2p$^1$P droplet excitation band. In rough agreement with earlier electron impact measurements~\cite{Callicoatt:1998}, the relative Ar$_{2}^{+}$ abundance falls off steeply as $N$ increases from $500$ to $4000$. Surprisingly, the relative abundance of alkali ions produced by Penning ionization follows the opposite trend. The relative abundance rises to a maximum around $N=4500$ for all species before falling off at even larger $N$.

%(*Check: What's the ratio of ionization at $h\nu <25\,$eV of alkali ionization versus Ar ?*)
%(*Charge transfer inwards, Penning outwards*)

Clearly, the simple picture of competing relaxation channels that limit the dopant ionization efficiency as adopted for charge transfer ionization of embedded species cannot be applied to Penning ionization of alkalis by excited He. On the contrary, the mobility of He$^*$ excitations appears to be enhanced in He droplets with intermediate size $N\sim 4500$, which disagrees with simple estimates that predict a shorter travel distance for He$^*$ than for He$^+$~\cite{Scheidemann:1993}. Possibly, the rising Penning signal when $N$ increases from 1000 to 5000 is related to the increased ratio of He atoms located in the bulk of the droplets with respect to those residing in the surface region where the He density $\rho_{He}$ is low ($60\%$ in surface region where $\rho_{He}<90\%$ of the bulk value for $N=400$ vs. $17\%$ for $N=4000$). In quantum wave packet simulations of the resonant charge migration, Seong~\textit{et al.} have found that in small He clusters ($N\leq 112$) large fractions of He$^+$ are trapped on single He sites or within small subsets of the He cluster~\cite{Seong:1998}, whereas in larger clusters delocalized states of He$^+$ are more likely to reach and ionize the dopant. This may be true a fortiori for the case of He$^*$ migration given the weaker He$^*$-He interactions and the resulting lower hopping speeds than for He$^+$ by a factor 2-3~\cite{Scheidemann:1993}. In addition to the delocalization of He$^*$ over a part of the droplet~\cite{Closser:2010}, attractive alkali-He$^*$ polarization forces may steer the He$^*$ migration towards the dopant just like He$^+$ charges are steered by long-range forces~\cite{Lewis:2005}. Eventually, when the He droplets exceed a certain size, the probability of the He$^*$ to reach the alkali impurity is likely to drop when the migration time exceeds the relaxation time into more deeply bound levels of He$^*$ (1s2s$^1$S, $^3$S) or He$_2^*$. Unfortunately, the dynamics of He$^*$ excitations in nanodroplets has not been studied in such detail as the dynamics of He$^+$ positive holes neither experimentally nor theoretically. Therefore no definite conclusions as to the droplet size dependence of the Penning process can be drawn at this stage.

\subsection{Photoelectron spectra}
The high efficiency of ionizing alkali atoms by energy transfer from He$^*$ as compared to direct photoionization may be related to the recently proposed optical dipole-like or resonant two-center ionization process~\cite{Najjari:2010}. In particular in the presence of $n\sim 10$ neighboring He atoms the ionization of an alkali atom via interatomic correlations may be collectively enhanced, resulting in the ratio of interatomic versus direct photoionization $n^2c^6/(\omega R)^6\sim 10^9$~\cite{Mueller:2011}. Here, $\omega=21.6\,$eV$/\hbar$ denotes the transition energy, $R\approx 6\,$\AA\, is the alkali-He distance and $c$ stands for the speed of light.

\begin{figure}[H]
\centering
\includegraphics[width=0.8\textwidth]{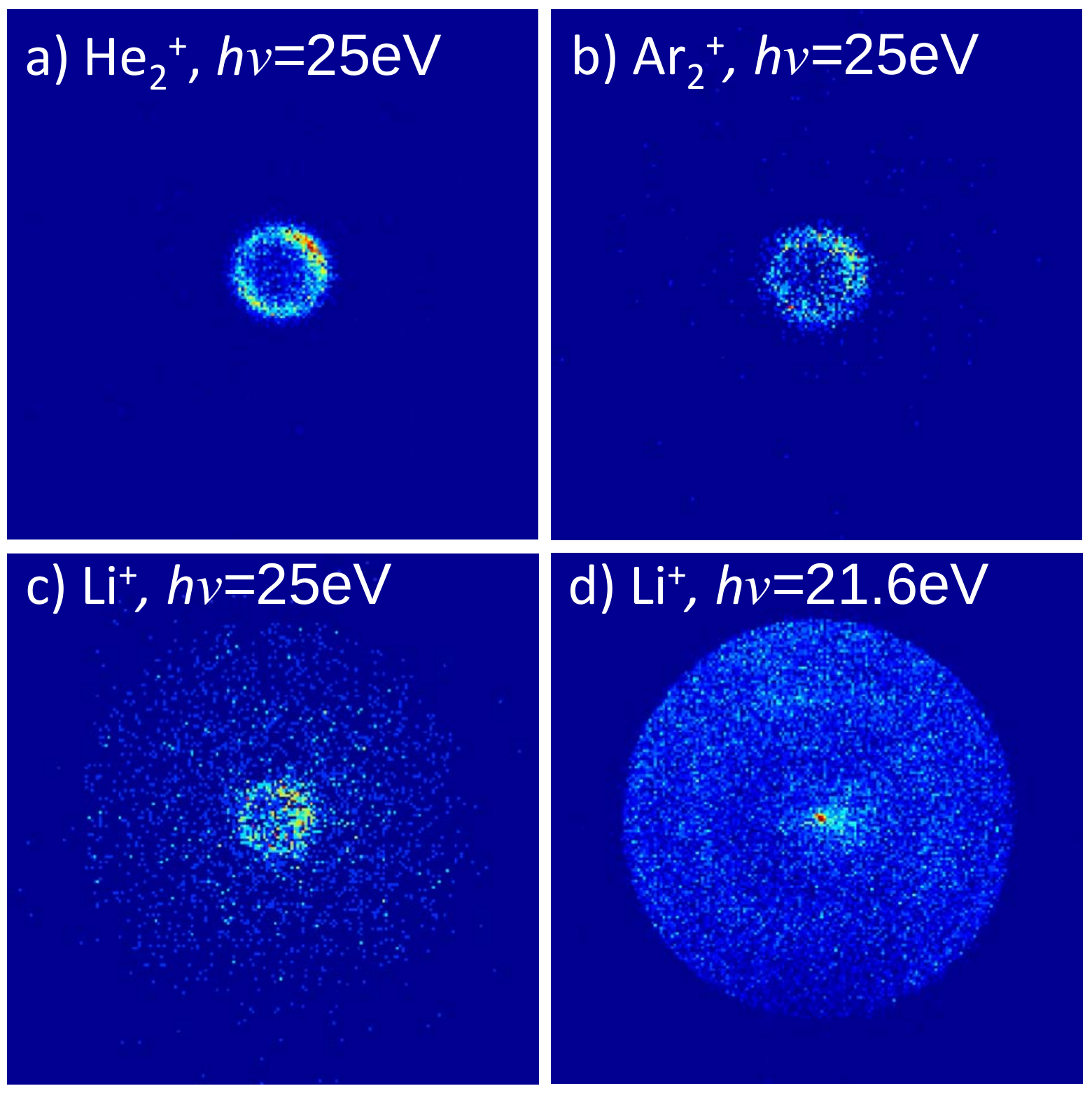}
\caption{Velocity-map images of photoelectrons measured in coincidence with He$_2^+$ (a), Ar$_2^+$ (b) and Li$^+$ ions at photon energies $h\nu=25\,$eV a)-c) and at $h\nu=21.6\,$eV d).}
\label{fig:VMI}
\end{figure}
In order to obtain more detailed information about the He droplet mediated ionization processes of dopants we analyze ion mass-correlated velocity-map images of photoelectrons. Fig.~\ref{fig:VMI} a) and b) show typical raw electron images recorded at photon energies $h\nu=25\,$eV in coincidence with He$_2^+$, Ar$_2^+$ and Li$^+$ ions, respectively. The electron image correlating to Li$^+$ shown in d) is recorded at $h\nu=21.6\,$eV. Note that the voltage settings of the imaging spectrometer used for a) and b) slightly differ from those used for c) and d). The images correlating to He$_2^+$ and Ar$_2^+$ show very similar small ring-shaped distributions, which confirms that Ar$_2^+$ ions are entirely generated by charge transfer ionization from ionized He droplets.

The electron distribution correlating to Li$^+$ recorded at the same photon energy $h\nu=25\,$eV (Fig.~\ref{fig:VMI} c)) shows two components. The central small ring again coincides with the one of He$_2^+$-correlated electrons indicating a large contribution of Li$^+$ ions created by charge transfer. A broader circular electron distribution is visible which roughly matches the one measured at $h\nu=21.6\,$eV (Fig.~\ref{fig:VMI} d)). This component corresponds to electrons created by ionization of Li dopants by the relaxation of excited He droplet states in a Penning-like process, as detailed below. We interpret the observation of Penning ionization even at $h\nu>E_{i,\mathrm{He}}$ by the action of bound electron-droplet states below the conduction band edge which is shifted in energy above $E_{i,\mathrm{He}}$ by about 1.1\,eV~\cite{Buchenau:1991}. Note that EUV fluorescence out of relaxed $^4$He droplet states was observed when exciting the droplets at $h\nu-E_{i,\mathrm{He}}$ up to 1.35\,eV~\cite{Haeften:2005}. Unfortunately the electron yield is too low to infer reliable information about the electron energy spectrum of the broad component. In addition to this broad circular distribution, Fig.~\ref{fig:VMI} d) features an intense central spot which corresponds to electrons with nearly vanishing kinetic energy. Note that no electrons are measured at that photon energy in the absence of Li dopants or in the absence of the He droplet beam.

\begin{figure}[H]
\centering
\includegraphics[width=0.6\textwidth]{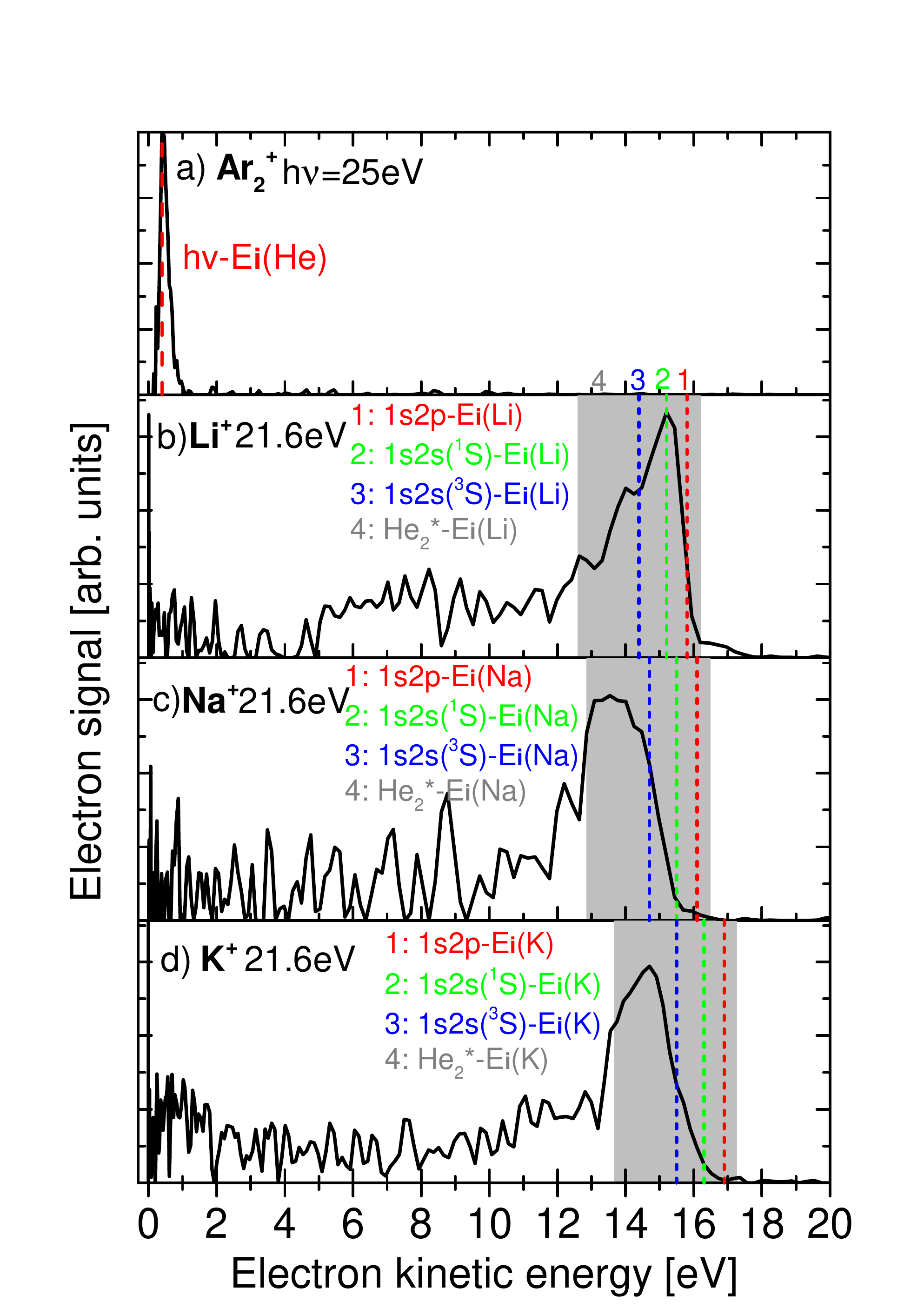}
\caption{Photoelectron spectra measured in coincidence with ions as indicated in panels a)-d). The Ar$_2^+$ spectrum shown in a) is recorded at $h\nu=25\,$eV, the alkali spectra b)-d) at $h\nu=21.6\,$eV. The dashed lines and the shaded area represent the expected photoelectron energies for charge transfer a) or Penning ionization b)-d) involving the directly excited 1s2p$^1$P state of He$*$ (1) or following relaxation into lower lying atomic (2,3) or molecular levels (4). The He expansion conditions are set to $p=50\,$bar and $T_0=19\,$K ($N=4500$).}
\label{fig:PEspectra}
\end{figure}
The photoelectron images are converted into photoelectron spectra as shown in Fig.~\ref{fig:PEspectra} by applying inverse Abel transformation and angular integration. Since the Ar$_2^+$ signal is produced entirely by charge transfer from ionized He, the photoelectron energy is peaked at $h\nu-E_{i,\mathrm{He}}=0.4\,$eV (dashed vertical line labeled by (1)). Unfortunately, no Ar$_2^+$-correlated photoelectron spectra with sufficient signal-to-noise ratio could be recorded at $h\nu=21.6\,$eV due to the much lower Penning ionization efficiency~\cite{Wang:2008}. The photoelectron spectra correlating to alkali metal ions at $h\nu=25\,$eV (not shown) closely resemble the one of Ar$_2^+$ as expected for charge transfer being the dominant ionization mechanism of the alkali dopants at $h\nu>E_{i,\mathrm{He}}$.

The photoelectron spectra in Fig.~\ref{fig:PEspectra} b)-d) are obtained from the integration of electron images recorded in coincidence with alkali ions at $h\nu=21.6\,$eV where excitation transfer ionization dominates. Note that the electron kinetic energy axis is carefully calibrated using the spectra of free He atoms recorded in a wide range of photon energies. The vertical dashed lines labeled (1)-(3) represent electron energies expected for transfer ionization out of various excited states of He$^*$, including the optically excited state 1s2p$^1$P labeled by (1) as well as the metastable lower lying states 1s2s$^1$S (2) and 1s2s$^3$S (3). The shaded area (4) indicates the range of energies of all possible vibronic levels of He$_2^*$~\cite{Buchenau:1991}. Our consideration of the unperturbed atomic He$^*$ and molecular He$_2^*$ levels is motivated by the fact that alkalis atoms are bound at a distance of 5-6\,{\AA} from the He droplet surface~\cite{Callegari:2011}, significantly more than the average distance between He atoms in the droplet interior of about 3.5\,\AA~\cite{Peterka:2007}. Thus we expect the He$^*$ excitation to be only weakly perturbed in the process of Penning ionizing the alkali dopant.

Despite the low resolution the spectra clearly indicate a shift of photoelectron energies to lower values with respect to the one expected for direct excitation transfer ionization out of the atomic 1s2p$^1$P state. Note that previous work on the direct photoionization of alkali dopants on He nanodroplets revealed a shift of the photoelectron energy by about 17\,meV toward higher energies due to the photoion-He interaction for which we find no indications~\cite{Loginov:2011}. In the Li case (Fig.~\ref{fig:PEspectra} b)), the photoelectron energy is peaked at that of the 1s2s$^1$S and 1s2s$^3$S levels, whereas for Na and K it seems that transfer ionization of the dopants is preceded by the formation of He$_2^*$ in various vibronic levels. This difference may be related to the larger delocalization of Li as compared to the other alkalis due to larger zero-point motion~\cite{Hernando:2011}.

The shifting of the Penning features with respect to the energy expected for direct ionization clearly points at a more complex, time-delayed ionization dynamics instead of the optical dipole-like process. Apparently this process involves the migration of the He$^*$ excitation through the He droplet, the relaxation into lower levels of He$^*$ and He$_2^*$, followed by ionization of the alkali dopant in a Penning collision-type process. Note that previous experiments with bulk superfluid He revealed the formation of He$_2^*$ in highly excited vibrational levels by tunneling through the He-He$^*$ barrier followed by slow relaxation into the ground state, as summarized in Ref.~\cite{Buchenau:1991} and references therein. This scenario of indirect Penning ionization is corroborated by the observation of a vanishing anisotropy parameter of the photoelectron angular distribution measured in coincidence with alkali ions $\beta=0.0(1)$, in contrast to $\beta=2.0(1)$ for direct photoionization of free He atoms and $\beta=0.8(1)$ measured in coincidence with He$_2^+$ or Ar$_2^+$ ions from doped He droplets at $h\nu=25\,$eV. If optical dipole-like ionization were the dominant process, both energies and angular distributions of photoelectrons correlated to the alkali ions would match those of directly ionized alkali atoms~\cite{Peterka:2006}, which is not the case.

Photoelectron spectra of He nanodroplets doped with Kr and Xe measured at $h\nu=21.6\,$eV showed that droplet-induced relaxation 1s2p$^1$P$\rightarrow$1s2s$^1$S is likely to precede Penning ionization of the dopants~\cite{Wang:2008}. However, no indications of further relaxation into lower-lying levels of He$^*$ or He$_2^*$ were found. The reason for the higher degree of relaxation observed in our experiments using alkali dopants may be related to longer interaction times of He$^*$ with the droplet environment prior to excitation transfer to the dopant. Since the He$^*$ excitations tend to migrate to the surface, that is away from the dissolved rare gas dopants Kr and Xe, the He$^*$-dopant interaction time is much shorter than in the case of alkali dopants. Further insight into the dynamics of migration and relaxation could be obtained from time resolved measurements using doped He nanodroplets similar to the ones recently performed with pure droplets~\cite{Kornilov:2010,Kornilov:2011,Buenermann:2012}.

\begin{figure}[H]
\centering
\includegraphics[width=0.6\textwidth]{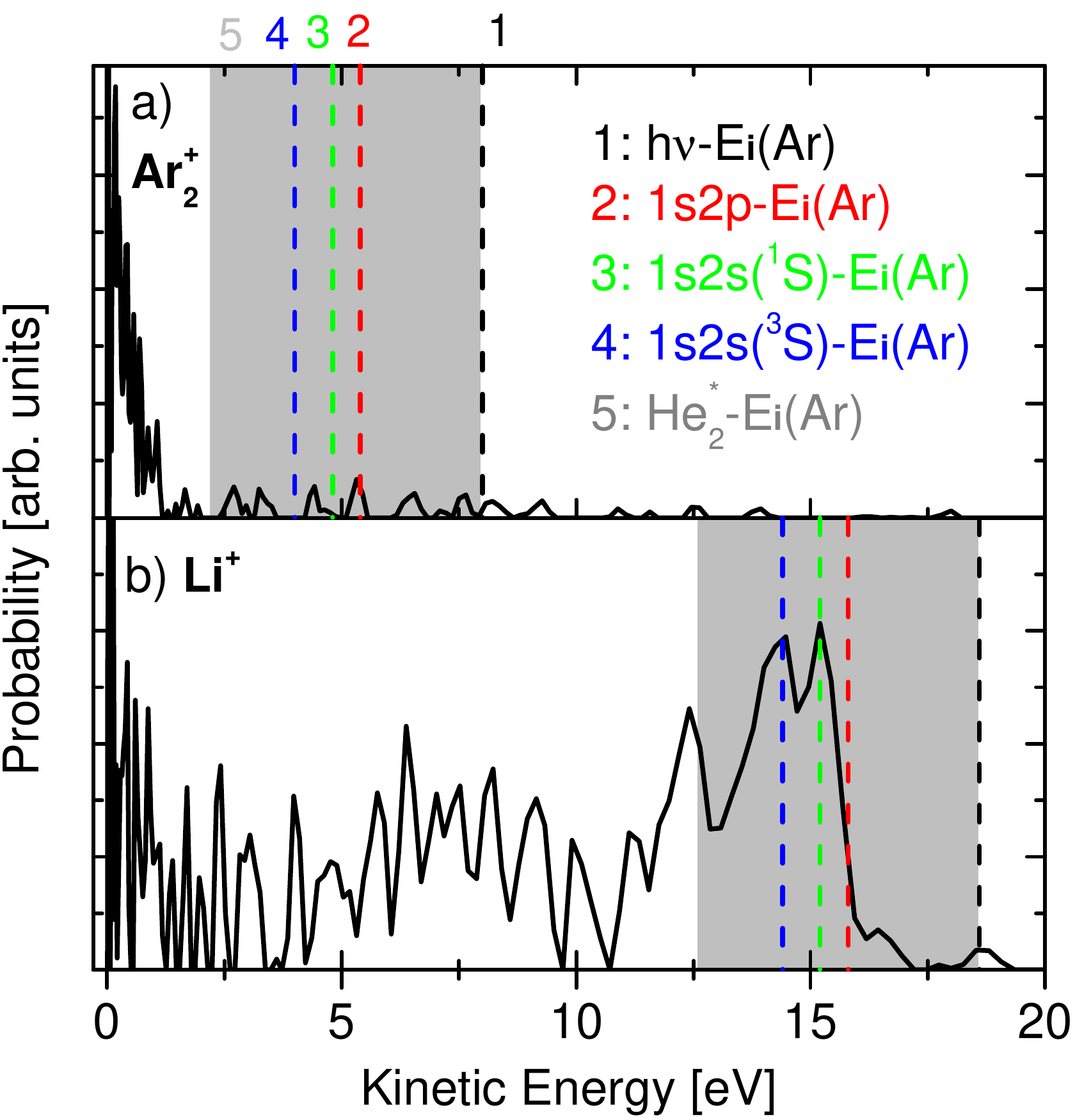}
\caption{Photoelectron spectra correlated to dopant ions Ar$_2^+$ measured at $h\nu=23.8\,$eV a) and to Li$^+$ at $h\nu=24\,$eV b). The vertical dashed lines in b) are labeled as in a) only that $E_{i,\mathrm{Ar}}$ is replaced by $E_{i,\mathrm{Li}}$.}
\label{fig:ZEKELiAr}
\end{figure}

Besides the sharp Penning features, the spectra contain significant signal contributions at low electron kinetic energies $\lesssim 12\,$eV visible as broad peaks in Fig.~\ref{fig:PEspectra} b) and Fig.~\ref{fig:ZEKELiAr} b) around 7\,eV and in Fig.~\ref{fig:PEspectra} d) around 1\,eV. Such broad spectral components were measured with better contrast in the photoelectron spectra of Kr and Xe-doped He droplets~\cite{Wang:2008}. In that work, Wang~\textit{et al.} observed a sharp increase of the low-energy feature for He droplet sizes ranging from $N=4000$ up to $250000$. These electrons, which are created by Penning ionization, appear to be subject to considerable loss of kinetic energy prior to ejection from the droplet. Our measurements show that the same mechanism of energy-loss is active also when Penning ionization occurs at the droplet surface.

Interestingly, the photoelectron spectra of dopants recorded at higher photon energies in the autoionization regime (ii) closely resemble those at $h\nu=21.6\,$eV, see Fig.~\ref{fig:ZEKELiAr}. The one correlating to Ar$_2^+$ recorded at $h\nu=23.8\,$eV shown in Fig.~\ref{fig:ZEKELiAr} a) again matches the one correlating to He$_2^+$. Since $h\nu<E_{i,\mathrm{He}}$ no direct ionization of He atoms is possible and only He$_2^+$ and electrons with very low kinetic energies are produced by autoionization~\cite{Peterka:2003,Peterka:2007}. This, however, implies that the He$_2^+$ charge remains mobile enough within the droplet to meet and ionize the Ar$_2$ dopant. Thus, we believe that the concept of resonant hopping of He$^+$ charges should be extended to He$_2^+$ and possibly to larger cluster ions which would involve He nuclear motion similar to the combined propagation of electronic excitation and diatomic proximity in an ensemble of Rydberg atoms~\cite{Wuester:2010}. The lack of photoelectron spectral components around $h\nu-E_{i,\mathrm{Ar}}=8\,$eV indicates that Penning ionization is again unlikely. This may appear unexpected given that the excited superposition of perturbed 1s3p$^1$P and 1s4p$^1$P states is delocalized over part of the He droplet~\cite{Closser:2010,Haeften:2011}. Apparently, the directed He$^*$ migration outwards to the surface nearly completely suppresses Penning ionization of dopants at the droplet center. Alternatively, autoionization followed by charge transfer ionization could occur when the He$^*$ excitation has just reached the Ar$_2$ dopant prior to Penning ionization. This scenario appears highly implausible, though, considering the large cross section of He$^*$-Ar Penning ionization collisions~\cite{Burdenski:1981} which even rises in the low collision energy regime provided by the cold droplet environment~\cite{Scheidemann:1997}.

The photoelectron spectrum recorded in coincidence with Li$^+$ at $h\nu=24\,$eV shown in Fig.~\ref{fig:ZEKELiAr} b) also closely resembles the one at $h\nu=21.6\,$eV. Only a small peak at an electron energy of 18.6\,eV indicates that a small fraction of Li dopants is ionized by excitation transfer from the originally excited droplet state whereas the larger part of the electron spectrum stems from Penning processes involving relaxed states of He$^*$ and He$_2^*$. Thus, the relaxation of 1s3p$^1$P and 1s4p$^1$P states proceeds at least as fast as relaxation of the 1s2p$^1$P droplet state.

\begin{figure}[H]
\centering
\includegraphics[width=0.6\textwidth]{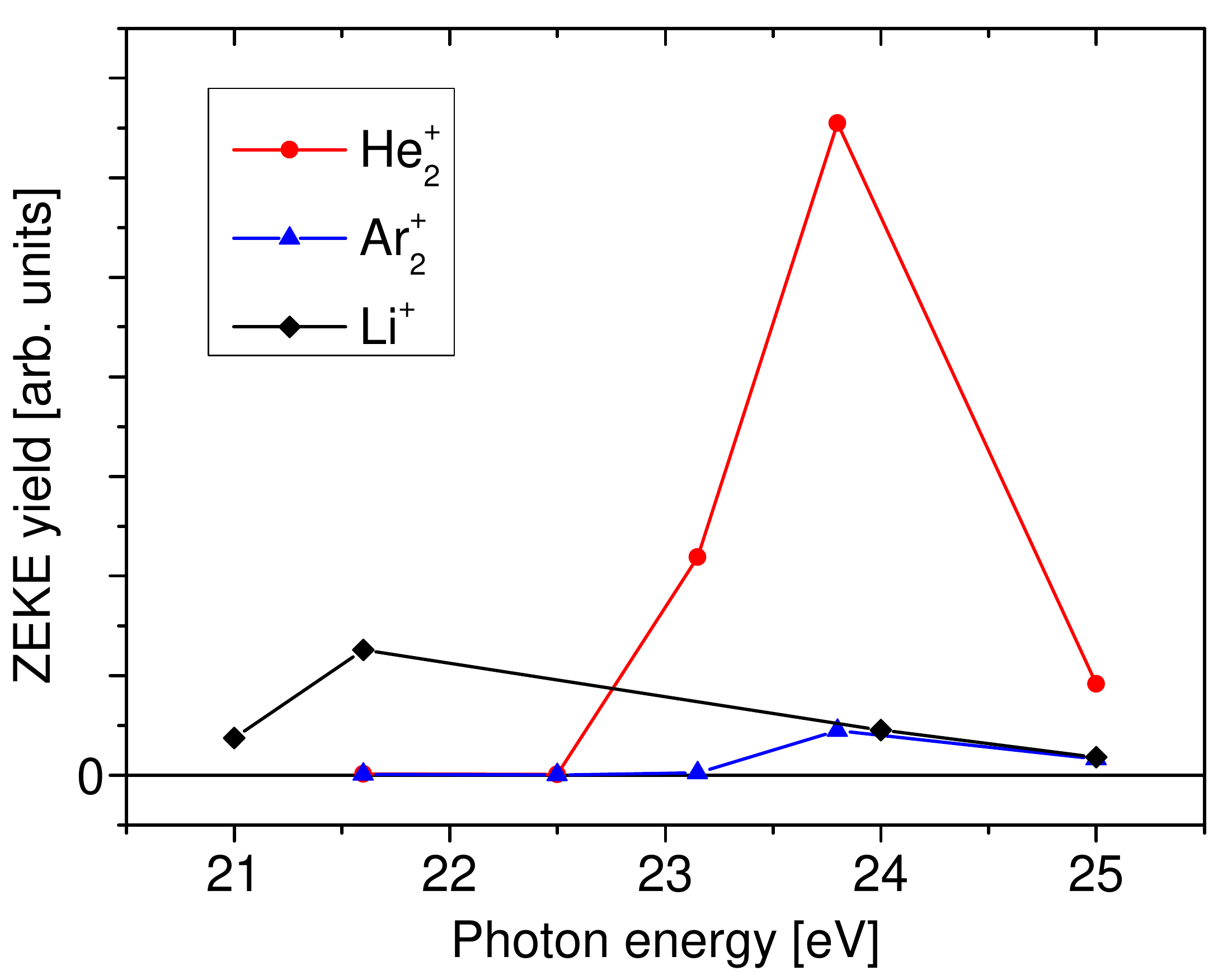}
\caption{Yield of electrons with very low kinetic energy (<10\,meV) correlating to the He$_2^+$, Ar$_2^+$, and Li$^+$ ions as various photon energies.}
\label{fig:ZEKE}
\end{figure}
Finally, we address the observation of photoelectrons with nearly zero kinetic energy (ZEKE) which appear in all experiments performed in regime (ii) of droplet autoionization (Fig.~\ref{fig:VMI} d)). While the exact formation mechanism of the ZEKE electrons is still under discussion, vibrational autoionization of highly excited states of the (doped) He droplet seems most likely~\cite{Peterka:2003,Peterka:2007,Wang:2008}. The appearance of ZEKE electrons has recently been measured in time-resolved experiments yielding a time constant of 1.5-2.5\,ps, limited by the dynamics of relaxation into lower-lying, non-emitting states~\cite{Kornilov:2010,Kornilov:2011}. We add to this discussion by measuring the ZEKE signal in coincidence with He$_2^+$, Ar$_2^+$ and Li$^+$ dopant ions. In order to avoid inaccuracies caused by artifacts from the inverse Abel transformation procedure, the ZEKE signal is determined by integrating the counts in the raw images within a small circular area with a radius corresponding to an electron energy of $10\,$meV.

The resulting ZEKE signal is shown in Fig.~\ref{fig:ZEKE} as a function of $h\nu$. Clearly, the highest ZEKE signal is measured in regime (ii) at $23 < h\nu < 25\,$eV in coincidence with He$_2^+$, which is the dominant product from autoionization and subsequent fragmentation of the He droplets. Note that the data point at $h\nu=25\,$eV highly overestimates the ZEKE signal since it mainly shows the contribution of photoelectrons with an energy of 0.4\,eV produced by direct He ionization which is partly included by the image analysis. The small ZEKE signal measured in coincidence with Ar$_2^+$ at $h\nu=23.8\,$eV results from autoionization of the He droplets and subsequent transfer ionization of argon, as discussed above.

Aside from ZEKE electrons produced by autoionization of the He droplets, we measure a considerable fraction of ZEKE electrons in coincidence with Li$^+$ dopants produced by Penning ionization at $h\nu=21.6\,$eV and at $h\nu=21\,$eV. A similar observation was previously made using He droplets doped with rare gases~\cite{Wang:2008}. Note that ZEKE electrons have also been seen for the case of directly laser-ionizing alkali dopants at excess energies of about 1.8\,eV~\cite{Fechner:2012}. The present observation of ZEKE electrons from an ionization process that involves an excess energy of up to 16\,eV that is dissipated by the He droplet clearly shows that a different mechanism of slowing down electrons presumably by collisions with He has to be active here which still eludes any founded interpretation.

\section{Conclusion}
We have reported synchrotron experiments to study the ionization dynamics of He nanodroplets doped with alkali metal atoms (Li, Na, K) which we compare to rare gas (Ar) dopants. In contrast to the rare gas dopants which are embedded in the droplet interior, surface-bound alkali atoms are found to be most efficiently ionized by the transfer of excitation from He$^*$ and He$_2^*$ in various excited states in a Penning-type process. We rationalize the enhanced Penning ionization rate by the directed migration of the excitation from the bulk outwards to the alkali dopants followed by excitation transfer. This contrasts the well-established hopping dynamics of a He$^+$ positive hole which is directed inwards into the droplet center where charge transfer ionization of embedded dopants occurs. However, the different dependencies of ionization efficiencies on the droplet size as well as photoelectron spectra correlated to the dopant ions indicate more complex migration and relaxation dynamics in the case of Penning ionization compared to charge transfer ionization. Simulations taking into account the structure of the doped He droplets as well as many-body correlation effects are required for gaining a better understanding of the dynamic response of doped He nanodroplets to photoexcitations.

\section{Acknowledgments}
The authors gratefully acknowledge support by the staff of Elettra for providing high quality light and P. Piseri and C. Grazioli for technical assistance with the PEPICO imaging detector. Furthermore, we thank O. Kornilov and K. von Haeften for stimulating discussions and the Deutsche Forschungsgemeinschaft and the Swiss National Science Foundation (Grant no: 200020\_140396) for financial support.

%\bibliography{achemso,RbHeBib}

\providecommand*\mcitethebibliography{\thebibliography}
\csname @ifundefined\endcsname{endmcitethebibliography}
  {\let\endmcitethebibliography\endthebibliography}{}

\end{document}